\documentstyle[aps,preprint]{revtex}

\begin{document}
\draft
\preprint{manuscript}
\title{Circulating and persistent currents induced by a current magnification 
and Aharonov-Casher phase}
\author{Taeseung Choi$^{a}$, Chang-Mo Ryu$^{a}$, 
and A. M. Jayannavar$^{b}$}
\address{$^{a}$ Department of Physics,
Pohang University of Science and Technology,
         Pohang 790-784, South Korea \\
$^{b}$ Institute of Physics, Sachivalaya Marg,
         Bhubaneswar-751005, India
         }
\date{\today}
\maketitle

\begin{abstract}
We considered the circulating current induced by the current 
magnification and the persistent current induced by Aharonov-Casher flux. 
The persistent currents have directional dependence on the direct current flow,
but the circulating currents have no directional dependence.
Hence in the equilibrium, only the persistent current can survives on the ring.
For the charge current, the persistent charge current cancelled between spin up
and down states, because of the time reversal symmetry of the Hamiltonian on the ring. 
So there are only circulating charge currents on the ring for electrons
with unpolarized spin in the nonequilibrium.
However, only the persistent spin currents contributes to the spin currents
for electrons with unpolarized spin.

\end{abstract}

\pacs{Keywords:
Mesoscopic systems, Ballistic transport, Spin-orbit coupling
}

\narrowtext
Recent studies in mesoscopic systems, over the entire sample of which quantum
coherence prevails, have provided several often counter-intuitive
new results. In mesoscopic samples at low temperatures the transport of 
quasiparticle is phase coherent and as a consequence several novel quantum
effects have been observed beyond atomic realm\cite{Kram}.
Existence of thermodynamic equlibrium presistent currents in mesoscopic rings 
is a manifestation of the Aharonov-Bohm effect which is being studied
intensively \cite{Chand,Mail,Cheu,Entin}.

Theoretical treatments up to date have been mostly concentrated on
isolated rings. Persistent currents occur not only in isolated rings but
also in the rings connected via leads to electron reservoirs, namely
open systems \cite{Butt1,Jay1}.
In a recent experiment Maily et al. have measured the persistent 
currents in both closed and open rings \cite{Mail}.
Recently Jayannavar et al. have noted the several novel effects related to
persistent currents can arise in open systems, which have no analogue in
closed or isolated systems \cite{Jay2,Jay4}.
Especially the directional dependence of persistent current in open 
system can be useful for separating the persistent current from noise.

As a dual of Ahraronov-Bohm phase, Aharonov and Casher (AC) \cite{b4} 
discovered the AC phase for a neutral magnetic moment encircling a charged line.
Aharonov and Anandan \cite{Anan} defined the nonadiabatic geometric phase for the cyclic 
evolution, called the AA phase, as a generalization of Berry's idea \cite{Berry}.
Qian and Su \cite{b13} has demonstrated the existence of
the AA phase in the AC effect.
Balatsky and Altshuler noticed
spin-orbit interaction produces persistent spin and mass currents 
\cite{Bal}.
The transport behavior  induced by the AC phase is recently 
studied \cite{Qia,Ours}.
And in our previous work \cite{pre}, we noticed that the directional dependence
of the spin currents induced by the AC phase.
We will generalize the system to the ring with different arm lengths.

Our system is depicted in Fig. 1. The Hamiltonian of our system is the 
same as that in our previous work \cite{pre}, i.e.,
\begin{equation}
H = \frac{1}{2m_e} ({\bf p} - \frac{\mu}{c} \mbox{\boldmath $\sigma$} 
\times {\bf E} )^2 + V \delta({\bf x} - {\bf x}_I),
\end{equation}
where $\mbox{\boldmath $\sigma$} \times \frac{{\bf E}}{2}$ represents 
a spin-orbit coupling, $\sigma^{\alpha}$ with $\alpha =1, 2, 3$ are 
Pauli matrices, and ${\bf x}_I$ is the position of the impurity.
In a closed ring, adopting a cylindrical coordinate system and the
electric field ${\bf E} = E( \cos \chi \hat{r} - \sin \chi \hat{z})$
we have the following Hamiltonian

\begin{equation}
H= \frac{\hbar^2}{2m_e a^2} \left(-i \partial_{\phi} -
\frac{\mu E a}{2 \hbar c}
(\sin{\chi} \cos{\phi} \sigma_x + \sin{\chi} \sin{\phi} \sigma_y
+\cos{\chi} \sigma_z)\right)^2,
\end{equation}
where $a$ is the radius of the ring.
The eigenfunctions $\Psi_{n, \pm}$ and eigenvalues $E_{n, \pm}$ of
Hamiltonian (2) in a closed ring are obtained as \cite{b9}
\begin{eqnarray}
\Psi_{n, \pm}& =& \frac{1}{\sqrt{2 \pi}} e^{in \phi}
                \left( \begin{array}{c}
                \cos{\frac{\beta_{\pm}}{2}} \\
                \pm e^{i \phi}\sin{\frac{\beta_{\pm}}{2}}
                \end{array} \right) ~, \nonumber \\
E_{n,\pm}&=&\frac{\hbar^2}{2ma^2} \left(n -
           \frac{\Phi^{\pm}_{\rm AC}}{2 \pi}\right)^2 ~,
\\ \mbox{and}\hspace*{1.5cm}
\Phi^{\pm}_{\rm AC}& =& - \pi(1 - \lambda_{\pm})~,\nonumber
\end{eqnarray}
where $\lambda_{\pm} \equiv \pm \sqrt{\omega_1^2 + ( \omega_3 + 1)^2}$
are eigenvalues of $\omega_1 \sigma^1 + (\omega_3 + 1) \sigma^3$, and
the angle $\beta_{\pm}$ are defined by
$\tan{\beta_{+}} \equiv \omega_1 / (\omega_3 +1)$, and
$\beta_{-} = \pi - \beta_{+}$. Here $\omega_1$ and $\omega_3$ are
denoted by $\omega_1 \equiv \frac{\mu E a}{\hbar c} \sin \chi$ and
$\omega_3 \equiv \frac{\mu E a}{\hbar c} \cos \chi$ and
$\mu = e \hbar / 2m_e c$ is the Bohr magneton.
The evolution of a spin state in the presence of the electric field is
determined by the following parallel transporter \cite{b9}.
\begin{equation}
 \Omega(\phi)
 = P\exp \Bigg[ i \frac{\mu E a}{2 \hbar c}\int_0^{\phi}
  (\sin{\chi} \cos{\phi'} \sigma^1 + \sin{\chi} \sin{\phi'} \sigma^2
  + \cos{\chi} \sigma^3) d \phi' \Bigg]  ,
\end{equation}
where $P$ is the path ordering operator.
It relates the wave function $\Psi(\phi)$ to $\Psi(0)$.

In our previous work, we considered the spin persistent currents of 
the open system where the lengths of two arms of the ring are same.
In that case, the spin persistent currents are induced by the
Aharonov-Casher (AC) phase.  
The AC phase induces the opposite direction of the persistent charge currents
for between the spin-up and the spin-down electrons since the Hamiltonian (1) 
has the time reversal symmetry. 
So if the  incident electrons are not polarized, the 
AC phase induces no net persistent charge currents. 
On the other hand, there are net spin currents since the spin operator
gets an additional minus sign under the time reversal operation.
But in the case the lengths of two arms of the ring are different, the
other important property occurs.
One of the authors (AMJ) and his coworkers showed that in the presence
of a current flow through the sample (the nonequilibrium situation), a net 
circulating charge current flows in a loop in the absence of external field 
in certain range of Fermi energy \cite{Jay2}.
First we will sketch how this net circulating current occurs in the open system
with different arms.
When one calculates the currents in two arms, there exists two possibilities
in general. In the first one, for a certain range of incident Fermi energies,
the currents in two arms are individually less than the total current.
In that case, the direction of the current through each arm will be the same
as that of the total current. In such a situation we do not assign a circulating
current on the ring. On the other hand, in a certain ranges of Fermi energies,
the magnitude of the current in one arm can exceed that of the total current
(current magnification). This implies that, to conserve the current, the direction 
of the current through the other arm must be opposite. In such a situation, one can
interpret the opposite current as a circulating current on the ring.
The magnitude of the circulating current is that of the opposite current.
This current magnification is the purely quantum mechanical property.
Very recently it has been shown that the same current magnification effect leads
to circulating thermoelectric currents highly exceeding the transport current
\cite{Mosk}.

In our system, there are two sources of rotating currents on the ring.
One is the current magnification and the other is the external flux 
(in our present case, the AC flux).
The current magnification occurs only in the nonequilibrium.
We divide the total current in the ring into the symmetric part
and the antisymmetric part with respect to the AC flux,
to understand the differnces of two sources clearly.
In this paper we will call the antisymmetric part as the persistent current following the denotation 
in our previous paper. 
In the nonequilibrium, for a certain range of incident Fermi energies, we can assign 
the circulating current to the symmetric part of the total current following the 
above paragrah.
Then the total rotating current on the ring is the sum of the persistent current and 
the circulating current.  
It depends on the spin direction like the persistent current.
In the previous case \cite{pre} there was no net persistent charge current 
for electrons with unpolarized spin even in the nonequilibrium situation,
since the arm lengths are equal to each other.

We first consider the case in which the current is injected from the left
reservoir ($\mu_L > \mu_R$, the nonequilibrium situation).
Where $\mu_L$ and $\mu_R$ are the chemical potentials of two electron reservoirs,
respectively. 
The lengths of the upper and the lower arms of the loop are $L_1$ and $L_2$, 
respectively. We have set the units $\hbar$, $2m$ to unity and all the lengths
are scaled with respect to the $L$ of the circumference of the loop 
($L= L_1 + L_2$).
At temperaure zero the transport current around a small energy interval $dE$ 
around $E$ is determined by $I = e T ( \mu_L - \mu_R) / 2 \pi$, where $T$
is the transmission coefficient of the system at the energy $E$.
To calculate the transmission coefficient $T$ and the currents in the upper
and the lower arms, we follow the our previous method of quantum waveguide transport
on networks \cite{pre,b15}. 
It is a straightforward exercise but somehow tedious to get the analytical expression.
And the resulting expression is too lengthy to express, so we will discuss our 
results graphically.

We have drawn the currents in Fig. 2 with the tilt angle $\chi = 2 \pi/3$, 
$kL=7.0 $, the impurity potential $V=2.0$, and $L_1 /L_2 = 5.0/3.0$ for varying 
the normalized field strength $\eta$ ($\equiv \mu E a/ \hbar c$).
We picked up the ratio of the arm lengths as the same as that in Ref. \cite{Jay2}.
In Fig. 2 the solid line shows the circulating charge current. One can readily
see that for small $\eta$ the nonadiabatic behaviors appear as we discussed in
our previous paper. 
And the direction of the current is only one direction, negative in Fig. 2.
For convenience we fix the positive of the flux as the direction going out of the 
paper in Fig. 1.  
And we will call the arm with the length $L_1$ as the upper arm and the other
as the lower arm.
Then for the counterclockwise rotating current, the direction of the rotating current 
in the upper arm is the opposite direction of the transport current to the right. 
Fowllowing the above convention, the circulating charge currents flow in the
clockwise direction only in Fig. 2.
And the maximum of circulating charge currents appears near the minimum in the
total transport current through the system,
which is represented by the dotted line.
The difference of arm lengths makes an antiresonance not be exact, but just appear
as a minimum. The another local minimums of the transport current, which have more
round shape in Fig. 2, is not related to the original antiresonances. 
They represent the contributions of the second harmonics like those in Fig. 1 of
Ref. \cite{Ours}.
And due to the presence of the impurity potential the minimum points of the transport
current, which is related to the antiresonance points of the loop structure,
is not exactly the same as the maximum points of the circulating charge currents.
It is because the multiple scatterings with the impurity potential shifts
the antiresonance points.
And the total rotating charge currents of spin-up electrons are represented as the 
dashed line.
The total rotating charge currents have both directions, clockwise and counterclockwise
directions like the persistent charge current shown in Fig. 3.
We show the persistent charge currents in Fig. 3 for both directions of direct current 
flows ($\mu_L > \mu_R$ and $\mu_R < \mu_L$). 
In Fig. 3 we have used the same parameters as in Fig. 2. 
It shows the persistent charge current of the spin-up electrons is
equal in amplitude and opposite in direction to that of the spin-down electrons
for both directions of direct current flows.
It implies there is no net persistent charge current for electrons with unpolarized
spin in the system of different arms also.
We can understand these results by the semicalssical argument in Ref. \cite{Ours,pre}.
The AC phase is the sum of the geometric phase (Aharonov-Anandan phase) 
\cite{Anan} and the dynamical phase due to the spin-orbit (SO) interaction \cite{Bal}.
According to our simple intuitive picture \cite{Ours,pre}, the dynamical phase can
be understood as the effective Aharonov-Bohm phase from the effective spin dependent
magnetic vector potential,
${\bf A}_{\rm eff} = ({\mu}/{2 e}) (\mbox{\boldmath $\sigma$} \times {\bf E})$.
For spin-up electrons, ${\bf A}^{+}_{\rm eff}$ becomes $\Phi^{\rm eff}_{\rm AB}
\cdot {\hat \phi}/ (2 \pi a)$, where $\Phi^{\rm eff}_{\rm AB} =
(\pi e a^2 E) \cos(\beta_+ - \chi)/ (2 m_e c^2)$. 
For spin-down electrons the vector potential becomes $- {\bf A}^{+}_{eff}$. 
Hence the directions of the dynamical fluxes are opposite to each other between the 
spin-up and the spin-down electrons.
And the geometric phase is nothing but the $-1/2$ times the solid angle subtended
by the curve of spin precession with respect to the origin. These become
$2 \pi (1 - \cos{\beta_+})$ for spin-up electrons and $2 \pi (1 + \cos{\beta_+})$
for spin-down electrons respectively.
Hence the directions of the geometric flux are opposite to each other modulo 
$2 \pi$. The modulo $2 \pi$ does not give any physical effects to the interferences, 
i.e., to the currents. 
Hence the direction of the total AC flux is also reversed by the reversing of the spin. 
In our system the total charge current on the ring depends on the AC phase $\Phi_{AC}$,
$L_1$, $L_2$, $L_d$, the Fermi energy $kL$, and chemical potentials $\mu_1$
and $\mu_2$. That is, the total charge current is a function of all these parameters,
$I^{(spin)}_{tot} (\Phi_{AC}, L_1, L_2, L_d, kL, \mu_1, \mu_2)$.
But the spin information manifests itself only through the AC phase in our system.
We are interested in the dependence on the AC flux and have
divided the total charge current into the symmetric part and the antisymmetric part 
with respect to the AC phase, $\Phi_{AC}$. 
So we note $I^{spin}_{tot} (\Phi_{AC}, L_1, L_2, L_d, kL, \mu_1, \mu_2) $ as
simply $ I^{spin}_{tot} (\Phi_{AC})$.
Since the spin information manifests itself only through the AC phase in our system,
$I^{+}_{tot} (\Phi_{AC}) = I^{-}_{tot} (-
\Phi_{AC})$ and $I^{+}_{tot} (- \Phi_{AC}) = I^{-}_{tot} (\Phi_{AC})$.
Then the persistent charge current for spin-up electron becomes the negative
of the persistent charge current for spin-down electrons as follows,
$$
I^{+}_{pc} \equiv \frac{1}{2} ( I^{+}_{tot} (\Phi_{AC}) - I^{+}_{tot}(- \Phi_{AC}))
\\ 
= - \frac{1}{2} ( I^{-}_{tot} (\Phi_{AC}) - I^{-}_{tot} (\Phi_{AC}))
= - I^{-}_{pc}.
$$   
On the other hand, the symmetric part of the total charge current
is the same for both spin-up electrons and spin-down electrons.
It implies that even for electrons with unpolarized spin, there are net rotating
charge currents from the circulating charge currents in the nonequilibrium.
These charge currents contribute for the total orbital magnetic moment of the ring.
But the absolute magnitude of the circulating charge current does not depend on the 
direction of the direct current in Fig. 4. 
The circulating charge current in left transport is the negative
of that in right transport.
To understand the directional dependence, it is better to consider only one spin
direction, e.g., spin-up here. The directional dependence is closely related to the 
time reversal symmetry breaking.
The circulating charge current is a sum of the contributions of both a positive AC
flux and a negative AC flux by definition.
Hence the circulating charge current does not detect any differences between
$\Phi^+_{AC}$ and $- \Phi^+_{AC}$.
It only observes the differences of the arm lengths. 
And the impurity potential does not prefer any direction of transport also. 
Hence if the current 
magnification takes place on a longer arm in the right transport, the current
magnification also apperars on a longer arm in the left transport.
It results the circulating charge current does not depend on the direction of the
direct current.
For a fixed value of the Fermi energy the circulating charge currents
changes only a sign as we change the direction of the current flow.
But it is natural the amplitude of the circulating charge current fluctuate according
to the variation of the AC flux. It is because the change of the AC flux gives
the similar effects to the change of the Fermi energy.
The dependence is not exactly the same as that of the Fermi energy, since the 
variation of the Fermi energy affects the entire sample but the variation of the
AC flux affects only the electrons on the ring.
For the persistent charge current, the difference of the sign of the AC flux
makes the preference of the direction. Hence it has the directional dependence.

In summary, we have considered the rotating charge and spin currents arsing from the 
current magnification and induced by the AC flux.
We have divided the total current into the symmetric part and antisymmetric
part with respect to the AC flux, to understand the different origins.
Then these two charge currents show totally different behaviors for different spins 
and directions of the current flow.
The persistent current depends on the direction of the current flow
as in the previous case \cite{pre}.
But the circulating  current does not depend on the direction of the transport current.
For a fixed value of the Fermi energy the the circulating charge currents
changes sign as we change the direction of the current flow.
As a result, in equilibrium (for spin polarized incoming electrons) 
the net charge currents in the system are only the 
persistent charge currents. 
However, in the presence of the transport current (the nonequilibrium), the net
circulating charge current flows in the ring by the current magnification,
even for electrons with unpolarized spin.
On the other hand, for spin currents, only the persistent spin current gives a
net spin current for electrons with unpolarized spin, since the spin
operator gets an additional minus sign under the time reversal operation.

T. Choi acknowledges for the support of the ICTP for his visit during
which the part of the present work was done.
This work was supported in part by Korean Research Foundation,
POSTECH BSRI special fund, and KOSEF.

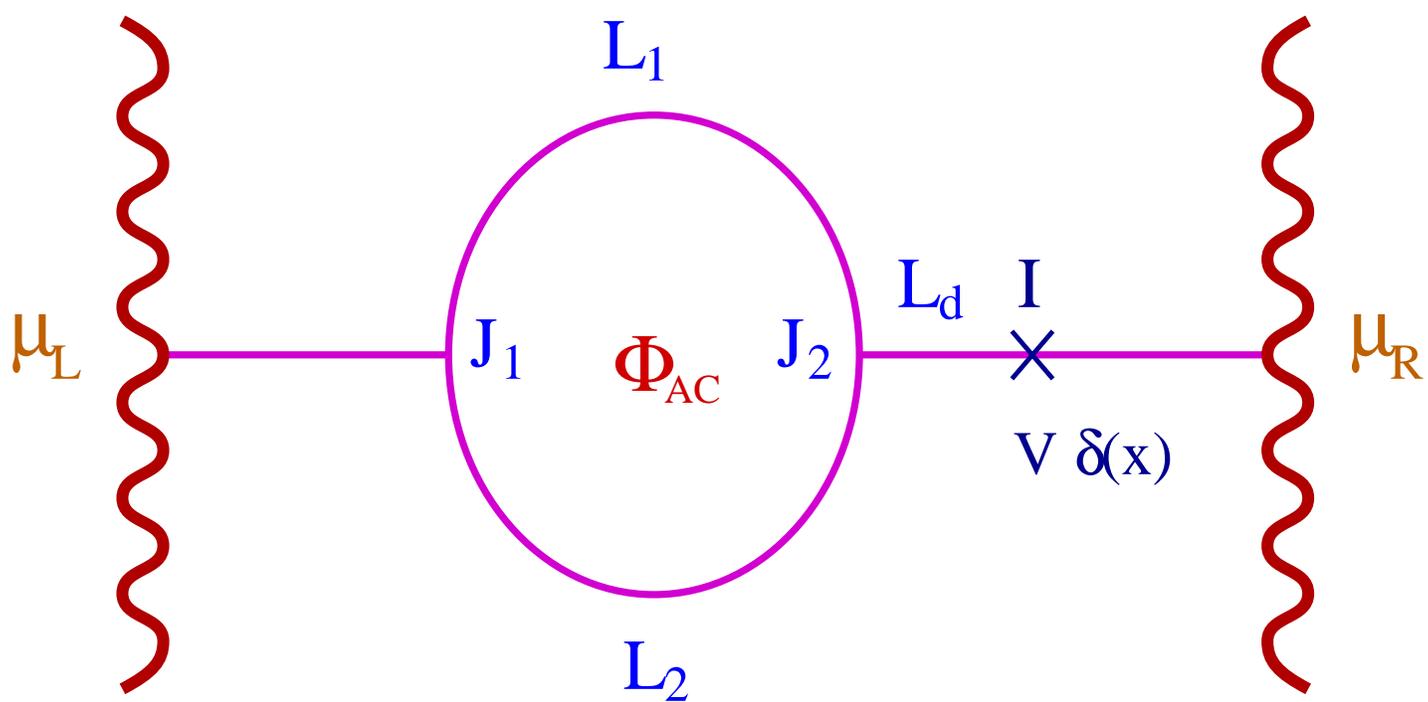
\begin{figure}
\caption{An open metallic loop connected to two electron reservoirs.
There exist a cylindrically symmetric electric field which gives
the AC flux.}
\label{fig1}\end{figure}

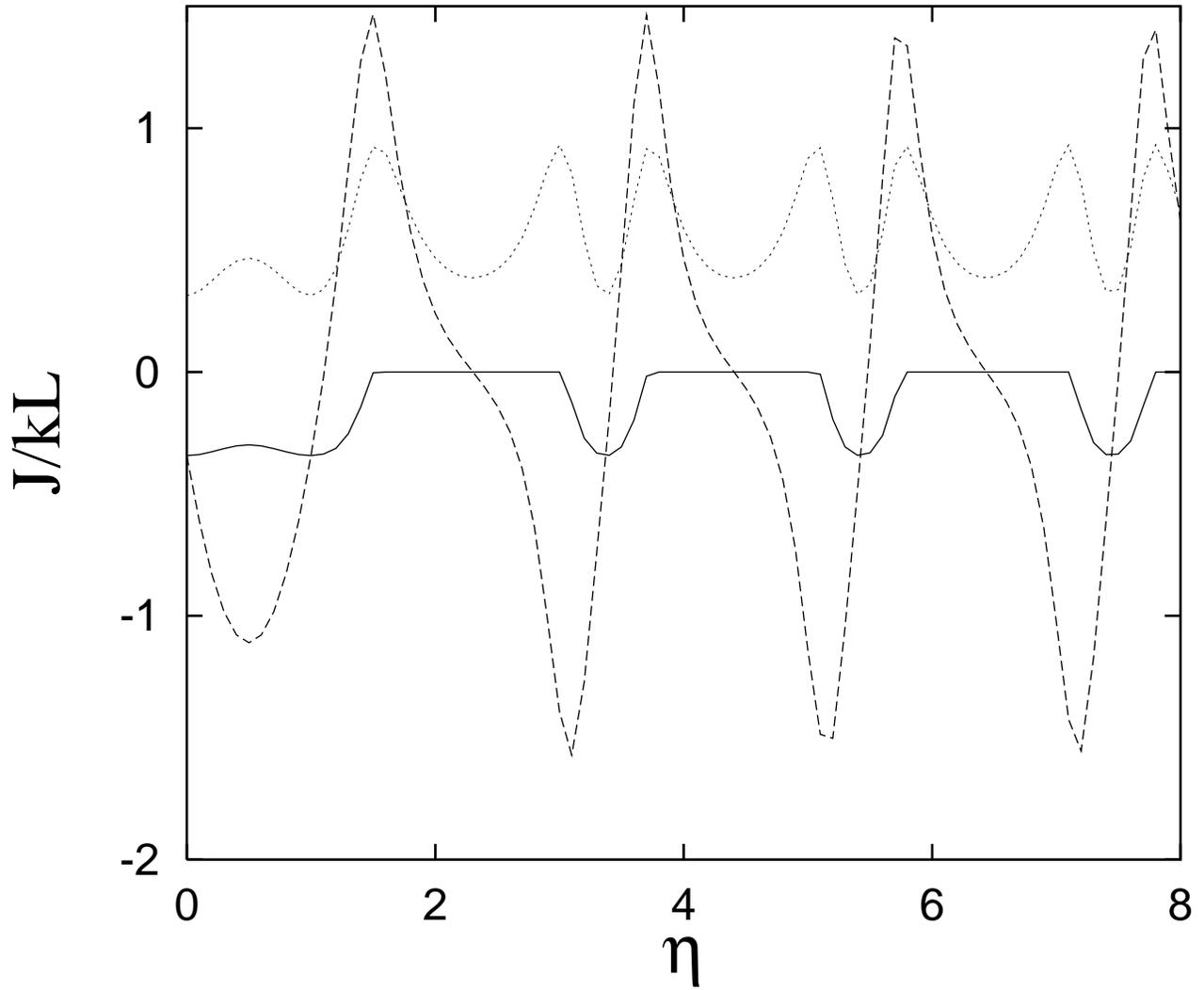
\begin{figure}
\caption{The charge currents as a function of the normalized
electric field $\eta$ for a fixed value of $kL=7$, $VL=2$, tilt angle
$\chi=2 \pi/3$, and $L_1 /L_2 = 5.0/3.0$.
The solid line represents circulating charge currents. 
The dotted line represents transport charge currents.
These two currents are the same for spin-up and spin-down electrons.
The dashed curve represents total rotating charge currents of spin-up electrons.
}
\label{fig2}\end{figure}

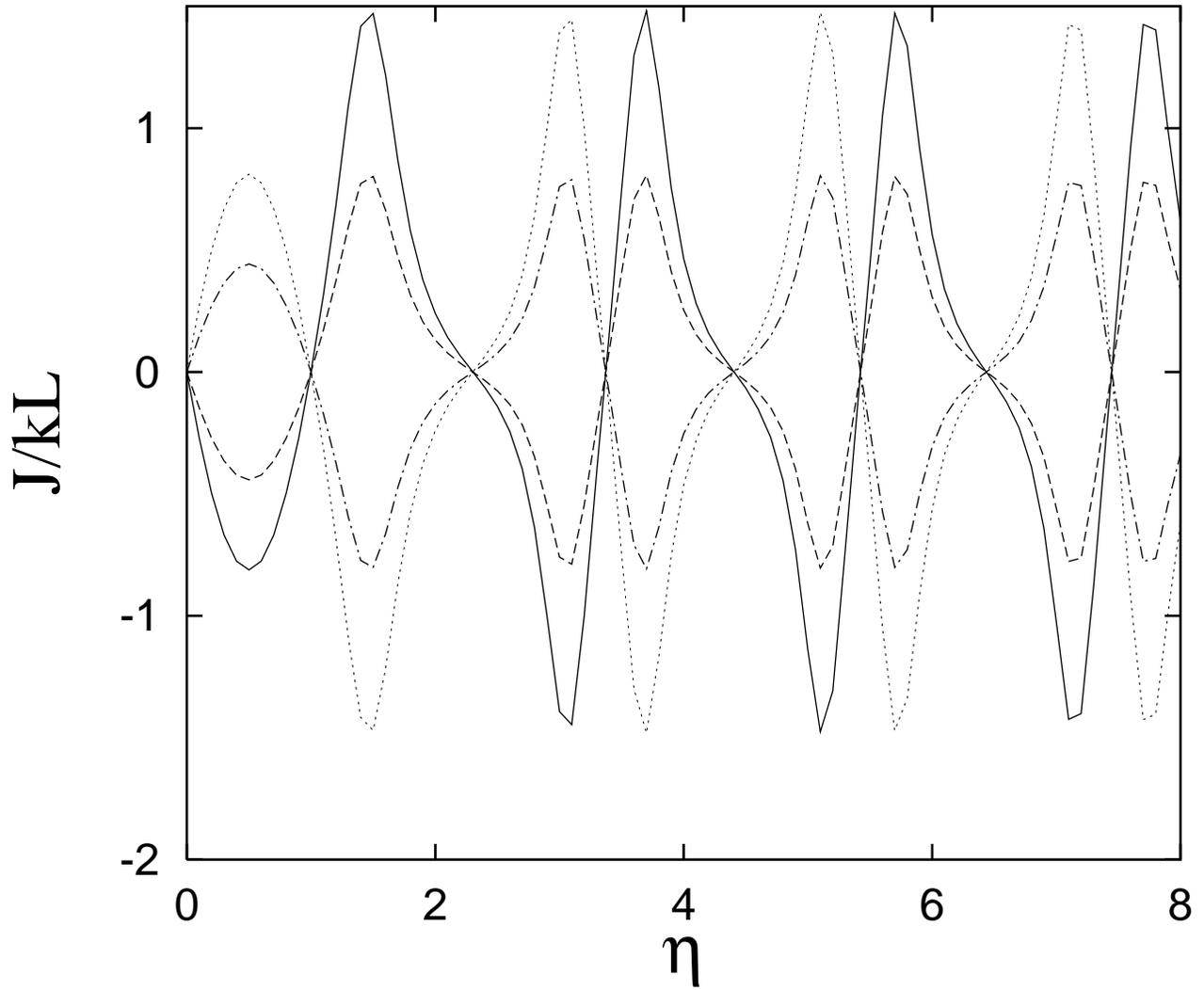
\begin{figure}
\caption{The persistent spin currents vs $\eta$ with the same parameters 
in Fig. 2.
The solid and dashed curves represents for the persistent charge current
of the electron with spin up eigenstate.
The dotted and dash-dotted curves are for spin down eigenstates.
This shows the cancellation between spin up and down persistent charge currents.
The persistent charge currents from left to right are greater than those
between spin up and spin down charge currents.
This shows the directional dependence.
}
\label{fig3}\end{figure}

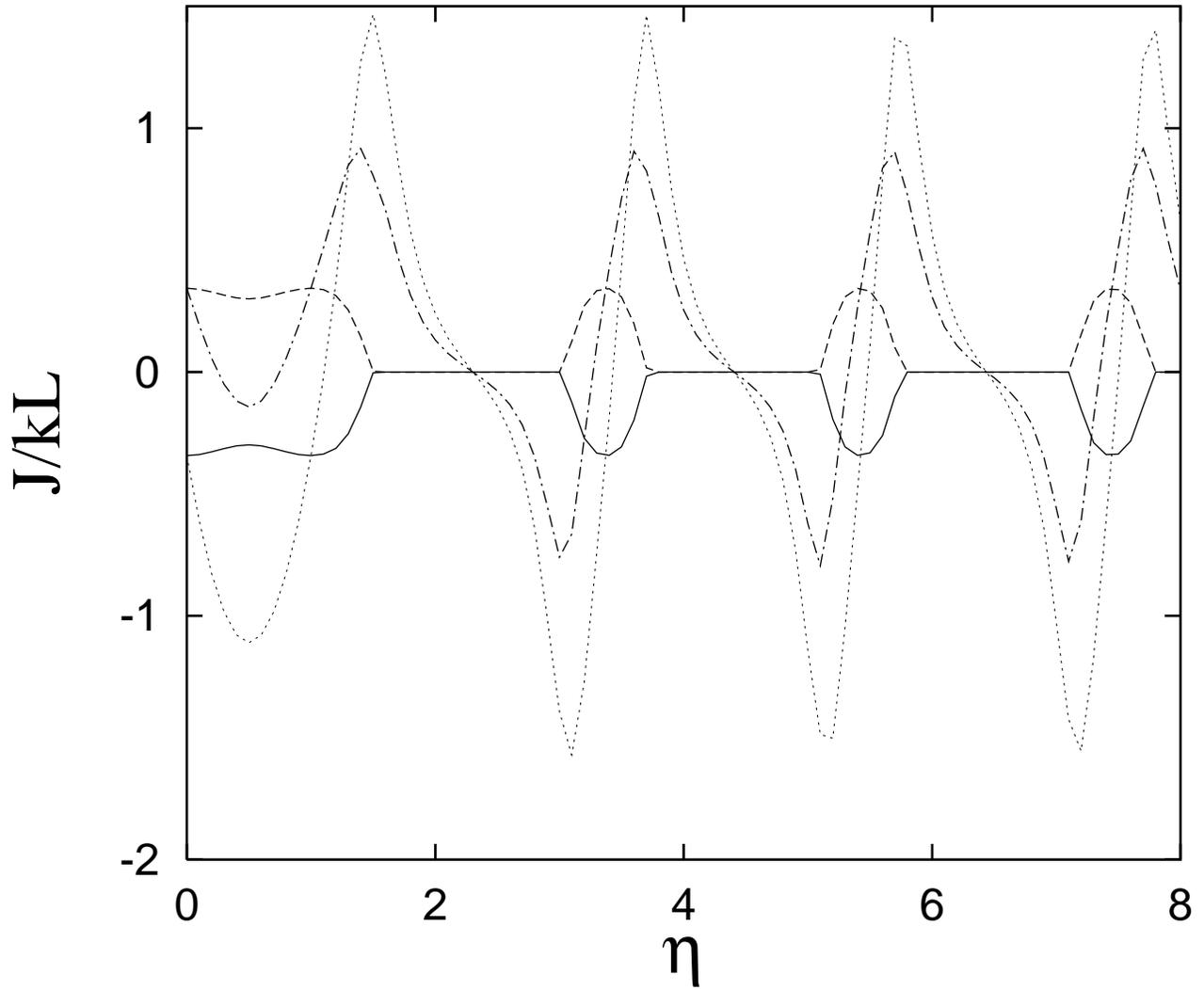
\begin{figure}
\caption{This shows directional dependence of the circulating current and the 
total rotating current for the same parameters against $\eta$ as other figures. 
The solid and dashed curves represents the circulating currents of left 
injected and right injected respectively.
And the dotted and dash-dotted represent the total rotating current of
left and right injected spin-up electrons respectively.
}
\label{fig4}\end{figure}

\end{document}